\documentclass[prl,aps,twocolumn]{revtex4}
\usepackage{epsfig}
\newcommand{\be}{\begin{equation}}
\newcommand{\ee}{\end{equation}}
\newcommand{\ba}{\begin{eqnarray}}
\newcommand{\ea}{\end{eqnarray}}
\newcommand{\ban}{\begin{eqnarray*}}
\newcommand{\ean}{\end{eqnarray*}}

\newcommand{\one}{\leavevmode\hbox{\small1\normalsize\kern-.33em1}}

\newcommand{\moy}[1]{\langle #1 \rangle}
\begin{document}
\date{\today}
\title{Tailoring photonic entanglement in high-dimensional Hilbert spaces}

\author{Hugues de Riedmatten, Ivan Marcikic, Valerio Scarani, Wolfgang Tittel, Hugo Zbinden and Nicolas
Gisin}

\affiliation{Group of Applied Physics,University of Geneva,20 rue
de l'Ecole-de-M\'edecine,CH-1211 Geneva 4,Switzerland }

\begin{abstract}
We present an experiment where two photonic systems of arbitrary
dimensions can be entangled. The method is based on spontaneous
parametric down conversion with trains of $d$ pump pulses with a
fixed phase relation, generated by a mode-locked laser. This leads
to a photon pair created in a coherent superposition of $d$
discrete emission times, given by the successive laser pulses.
Entanglement is shown by performing a two-photon interference
experiment and by observing the visibility of the interference
fringes increasing as a function of the dimension $d$. Factors
limiting the visibility, such as the presence of multiple pairs in
one train, are discussed.
\end{abstract}
\maketitle Entanglement is one of the essential features of
quantum physics. It leads to non classical correlation between
different particles. Entanglement of two-levels systems (qubits)
has been extensively studied, both theoretically and
experimentally, in order to perform fundamental tests of quantum
mechanics and to implement a number of protocols proposed in the
burgeoning field of quantum information science (see e.g.
\cite{tittel01} for a recent review). However, it is interesting
to explore higher-dimensional Hilbert spaces. From a fundamental
point of view, increasing the complexity of the systems and the
dimension of the Hilbert space might lead to a further insight
into the subtleties of quantum physics. For instance,
high-dimensional entangled states give experimental predictions
which differ more radically from classical physics
\cite{Kaszlikowski00,collins02} than entangled qubits. They could
also decrease the quantum efficiency required to close the
detection loophole in Bell experiments \cite{massar01}. In the
more applied context of quantum information science, high
dimensional entangled states might also be of interest. In
particular, high-dimensional systems can carry more information
than two-dimensional systems and increase the noise threshold that
quantum key distribution protocols can tolerate
\cite{hbp2000,cerf01}. Moreover, using entangled qudits might
increase the efficiency of Bell state measurements for quantum
teleportation \cite{Witthaut03}. Finally, although most of the
proposed protocols require only entangled qubits, some protocols
involving qutrits (3-dimensional systems) have been recently
proposed, such as the Byzantine agreement \cite{fitzi01} and
quantum coin tossing \cite{coin}.

Only recently the first experiments started to explore
entanglement in higher dimensions. Two directions can be
considered. First, one can take profit of multi-photon
entanglement, as obtained for example in higher order parametric
down conversion \cite{Lamas01,Howell02}. Second, one can use the
entanglement of two high-dimensional systems. Entanglement of
orbital angular momentum of photons has been for instance proposed
and demonstrated in this context \cite{Mair01,Vaziri02}.
Energy-time entanglement has also been recently analyzed in 3
dimensions \cite{thew03Q}, using unbalanced 3-arm fiber optic
interferometers in a scheme analogous to the Franson
interferometric arrangement for qubits.

All these methods so far have been demonstrated only for qutrits
and it will be difficult to implement them in higher dimensions.
In contrast, we recently proposed a simple method to entangle two
photonic systems of arbitrary dimensions. It is based on
spontaneous parametric down-conversion (SPDC) with a sequence of
pump pulses with a fixed phase relation generated by a mode-locked
laser, leading to high-dimensional time-bin entanglement
\cite{HdR02}. In this paper, we report on an experimental
realization of this scheme, where it is possible to choose
arbitrarily the dimension of the entangled photons Hilbert space.
An advantage of our scheme is that it enables the generation of
entangled states in arbitrary dimensions in a scalable way with
only two photons \cite{note0}. We perform a simple analysis which
is sufficient to show entanglement, although it does not provide a
full information about the states.

Before describing the experiment, let us recall the basics of
high-dimensional time-bin entanglement. Suppose a SPDC process
with a train of $d$ pump pulses with a fixed phase relation.
Providing that the probability of creating more than one pair in
$d$ pulses is negligible and excluding the vacuum, the state after
SPDC is \cite{HdR02}:
\begin{eqnarray}
|\Psi\rangle_{PDC} = \sum_{j=1}^{d}c_je^{i\phi_j}|j,j\rangle
\label{psiPDC}
\end{eqnarray}
where $|j,j\rangle \equiv |j_A,j_B\rangle$  corresponds to a
photon pair created by the pulse (or time-bin) j, with relative
amplitude $c_j$ and phase $\phi_j$. The phase reference $\phi_1$
is set at 0. $A$ and $B$ are the two SPDC modes, $d$ is an integer
that can be arbitrarily large and  $ \sum_{j=1}^{d}c_j^2=1$.

This method enables to create any bipartite high-dimensional
state. By selecting the number of pump pulses we can choose the
dimension of the entangled photons Hilbert space. In our
experiment we construct trains of $d$ pulses, where $d$ can be
varied from 1 to 20, with constant amplitudes $c_j$ and with
constant phase shifts $\phi_j-\phi_{j-1}=\phi=cst.$ Note that by
inserting a phase and/or amplitude modulator before the
down-converter, we could in principle modulate their amplitudes
and phases, thus varying the coefficients $c_j$ and $\phi_j$ in
order to generate arbitrary non-maximally entangled states.
\begin{figure}[h]
\includegraphics[width=8 cm]{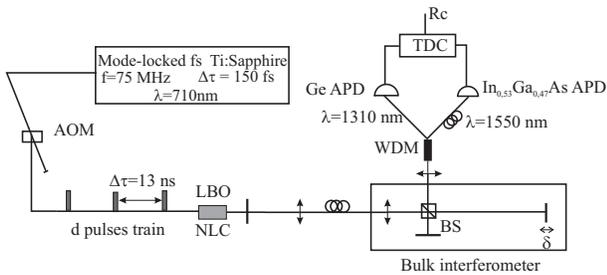}
\caption{Schematic of the experiment. See text for details.}
\label{setup}
\end{figure}

A complete analysis of such high dimensional entangled states
would require the use of d-arm interferometers, such that the
amplitudes and phases of all time-bins can be probed. An
alternative could be given by the use of fiber loops or
Fabry-Perot interferometers, as proposed in \cite{HdR02}. Here, we
used a 2-arm interferometer, which already shows high-dimensional
entanglement. The travel time difference between the long and the
short arm of this interferometer is equal to the time between 2
pump pulses $\Delta \tau$ (see Fig. \ref{setup}).
 This means that a photon travelling through
 the short arm will remain in the same time-bin while a photon travelling
 through the long arm will move to the next time-bin. We restrict ourself to the events
where both photons of one pair travel the same path in the
interferometer, and are thus detected with a time difference
$\Delta t=t_A-t_B=0$. In this case, the evolution of the state of
Eq.(\ref{psiPDC}) in the interferometer can be written as (not
normalized):
\begin{eqnarray}
\nonumber \left|\Psi_{int}\right\rangle &=&\left|1,1\right\rangle
+\sum_{j=2}^{d}\left|
j,j\right\rangle(e^{\phi_{j}}+e^{i(\delta_A+\delta_B+\phi_{j-1})})\\
&+&e^{i(\delta_A+\delta_B+\phi_d)}\left|d+1,d+1\right\rangle
\label{psiint}
\end{eqnarray}
where $\delta_{A,B}$ are the phases introduced in the long arm of
the interferometer for the photons $A$ and $B$ and with
$\phi_1=0$. We see that for all time-bins except the first and the
last one we have a superposition of two indistinguishable
processes. If we record all the processes leading to a coincidence
with $\Delta t=0$, i.e if we don't postselect the interfering
terms, the coincidence count rate varies as
\begin{equation}
R_{c}\sim 1+V_d\cos (\delta_A+\delta_B-\phi)
\end{equation}
where $\phi_{j}-\phi_{j-1}=\phi$ for all $j$. From the $2d$
different processes, two are always  completely distinguishable
(the first and the last time bin). Therefore, the maximal
visibility of the interference fringes, $V_d$, depends on the
dimension $d$ as:
\begin{equation}
V_d=V_{max} \frac{d-1}{d} \label{visD}
\end{equation}
where $V_{max}$ is the maximum visibility due to experimental
imperfections. This analysis is valid if the phase difference
between 2 pulses is constant, which is the case in a mode-locked
laser. Two contributions might affect the stability. First the
laser cavity length may vary slowly due to thermal drift. This
drift has been measured ($\sim 2 \mu m$ per hour) and is
negligible in the time-scale of a round trip time. Second, one
could imagine faster fluctuations of the optical cavity length due
e.g. to mechanical vibrations. This seems however unlikely, since
important fluctuations would destroy the laser operation. To
further confirm this point, we make the following reasoning. If we
consider a small phase noise between 2 consecutive pulses with a
Gaussian distribution of width $\delta\epsilon$, the visibility
will be reduced to: $V=V_d\exp(-\frac{1}{2}\delta\epsilon^2)$. The
phase noise between pulse j and pulse j+m also has a gaussian
distribution of width $\sqrt{m}\delta\epsilon$, leading to a
visibility $V=V_d\exp(-\frac{1}{2}(m)\delta\epsilon^2)$. Observing
a visibility $V_d$ close to optimal is thus a confirmation that
the phase noise $\delta\epsilon \ll \pi$ and consequently that the
coherence is maintained over many time bins.

In our experiment, we use trains of $d$ pump pulses, where $d$ can
be varied from 1 to 20, and we observe the visibility of the two
photon interference as a function of the dimension $d$. A
schematic of the experiment is presented in Fig. \ref{setup}. The
pump laser is a Ti-Sapphire femtosecond mode-locked laser
producing 150fs pulses at a wavelength of 710 nm. The time between
2 pulses is $\Delta \tau$ = 13 ns. To construct the pulse trains,
the pump beam is focussed into a 380MHz acousto-optic modulator
(AOM, from Brimrose) which reflects the incoming beam with an
efficiency of $\approx 50\%$ when it is activated. This activation
can be triggered externally, with a TTL signal of variable width
synchronized with the laser pulses. The rise time is around 6ns.
The width of this signal thus determines the number of pulses per
train. The reflected beam containing the pulse trains is then used
to pump a non-linear Lithium triborate (LBO) crystal. Non
degenerate photon pairs at 1310/1550 nm wavelength are created by
SPDC and then sent to the analyzer, which is a two-arm bulk
Michelson interferometer, where the long arm introduces a delay
$\Delta \tau$=13 ns with respect to the short one, corresponding
to a physical path-length difference of 1.95 m \cite{note1}. The
pump power is kept low, in order to keep the probability of having
more than one pair per train small. Photons exiting one output of
the interferometer together are first focussed into an optical
fiber and then separated with a wavelength division multiplexer
(WDM). The 1310 nm photon is detected by a passively quenched LN2
cooled Ge Avalanche Photo Diode (APD, from NEC), with a quantum
efficiency $\eta$ of around 10$\%$ for 40kHz of dark counts. The
1550 nm photon is detected with an InGaAs photon counting module
(from idQuantique), featuring a quantum efficiency of around
30$\%$ for a dark count probability of $10^{-4}$ per ns and
operating in gated mode. The trigger is given by a coincidence
between the Ge APD and a 1-ns signal delivered simultaneously with
each laser pulse ($t_0$), in order to reduce the accidental
coincidences. The signals from the APDs are finally sent to a
time-to-digital converter, in order to record the photons arrival
time histogram. A small coincidence window of around 1 ns is
selected around the interfering peak (i.e the peak with $\Delta
t=0$).

\begin{figure}[h]
\includegraphics[width=8cm]{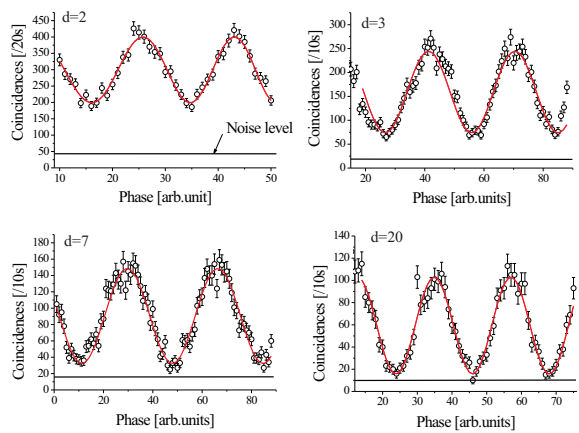}
\caption{Two-photon interference visibility for different
dimensions d. The solid line is a sinusoidal fit from which we can
deduce the net visibility of the fringes. The level of accidental
coincidence is indicated by the straight line.} \label{interfd}
\end{figure}

\begin{figure}[h]
\includegraphics[width=8cm]{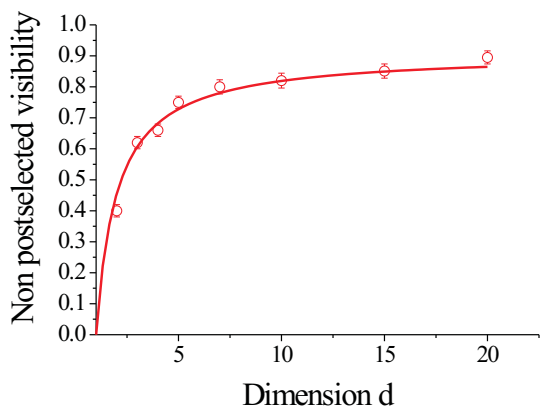}
\caption{Two-photon interference visibility as a function of the
dimension of the Hilbert space. The black circles are experimental
points. The solid line is a fit with eq. \ref{visD}}.
\label{results}
\end{figure}

If we record the coincidence count rate as a function of the phase
shift in the interferometer, we obtain
 sinusoidal curves with a visibility increasing with the
dimension $d$ (see Fig.\ref{interfd}). Net visibilities (i.e with
accidental coincidence count rate subtracted) as a function of the
dimension $d$ are plotted in Fig.\ref{results}. The solid line is
a fit using Eq. (\ref{visD}). The good agreement between
experimental data and theory confirms that the dimension of the
entangled photons is given by the number of pump pulses d. We find
a maximal visibility of $91.6\pm 1.2\%$.

We now discuss the factors limiting the visibility which, as we
will see, is not reduced by a possible phase noise between pump
pulses. A first factor is the possible creation of more than one
pair per pulse train. The spectral bandwidth of the (not filtered)
created photons is about 100nm, corresponding to a coherence time
of $\approx 25 fs$, much smaller than the duration of the pump
pulse. In this limit, any $2n$ photons state can be described as
$n$ independent pairs. The probability of producing $n$ pairs in a
given pulse is distributed according to the Poissonian
distribution of mean value $\mu$: $p_n=
e^{-\mu}\,\frac{\mu^n}{n!}.$ The starting point for the
calculation of the loss of visibility due to multiple pairs is the
fact that the total coincidence count rate can be written \ba
R&=&R_{1}\,(1+V_d\cos\theta)\,+\,R_2\,. \label{totrate}\ea The
first term of the sum means that, for each pair created, the
two-photon process described above can take place, leading to an
interference fringe of visibility $V_d$. The additional rate $R_2$
is what comes from the multi-pair pulses, when one detects
coincidence of photons belonging to independent pairs. In our case
there are only two kinds of contributions to $R_2$: either the
photons were created in the same time-bin ($R_{2,s}$), or in
consecutive time-bins ($R_{2,c}$); if the independent pairs are
created in more distant time-bins, no coincidence is registered.

\begin{figure}[h]
\includegraphics[width=8cm]{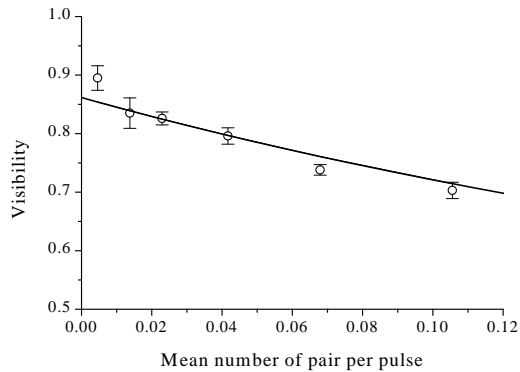}
\caption{Visibility of the interference fringes as function of the
mean number of pairs per pulse $\mu$, for d=20.} \label{vmu}
\end{figure}

Now, we calculate $R_1$, $R_{2,s}$ and $R_{2,c}$ explicitly. $R_1$
is proportional to the mean number of pairs created $\mu d$. The
factor of proportionality is given by the probability that a
photon pair leads to a coincident detection (i.e with $\Delta
t=0$), which is $\frac{1}{2}$ \cite{note2}. Hence finally $
R_1=\frac{1}{2}\,\mu \,d$. Let us now calculate $R_{2,s}$. With
$n$ pairs in a given time-bin, one can create $\frac{n(n-1)}{2}$
couples, so the mean number of such couples in d time-bins is
$d\sum_n p_n \frac{n(n-1)}{2}=d\frac{\mu^2}{2}$. By inserting the
probability of coincidence \cite{note4}, we find: $
R_{2,s}=\frac{\mu^2}{2} \, d\,. $ Let us finally calculate
$R_{2,c}$. If $n_k$ is the number of pairs in time-bin $k$, the
number of pairs in consecutive time-bins is $m=n_1\times
n_2+n_2\times n_3+...+n_{d-1}\times n_d$. The average of the
random variable $m$ is
$\moy{m}=\sum_{n_1}...\sum_{n_d}p_{n_1}...p_{n_d}
\,m(n_1,...,n_d)\,=\,(d-1)\,\mu^2\,$. In this case, only half of
the processes lead to a coincident detection. We thus obtain $
R_{2,c}=\frac{1}{2}\,\mu^2 \, (d-1)\,$. Inserting these results
into (\ref{totrate}) we find $R\,\propto\,1+V(\mu,d)\cos\theta$
with \ba V(\mu,d)&=&V_d\,\frac{1}{1+2\mu-\mu/d}\,. \label{Vp}\ea
To validate our model, we measured the visibility as a function of
$\mu$, for $d$=20 (see Fig. \ref{vmu}). The factor $\mu$, which is
proportional to the pump power is determined by the side peak
method, explained in detail in \cite{marcikic02}. The solid line
is a fit of Eq.(\ref{Vp}), in good agreement with the experimental
data.
 For the measurement of Fig. \ref{results} (not corrected), $\mu$ is kept low ($<0.025$) so that
 we estimate the maximal visibility due to multiple pairs to
 $(97 \pm 1)\%$.

Another factor that affects the visibility is the non perfect
alignment of the analyzing interferometer. Ideally, the
transmission in the long and the short arm should be the same for
both wavelength. Due to the fact that the interferometer is long
and that the two photons have different wavelengths, obtaining a
good alignment is very difficult. To calculate the influence of a
misalignment we write $t_s$ and $t_l$ the transmission probability
amplitudes for the short and the long arm respectively. For
simplicity, we assume them to be the same for both wavelengths. In
this case the coincidence count rate (if we take only the
interfering terms) is $R_{c}\sim t_s^4+t_l^4+2t_s^2t_l^2\cos
(\delta_A+\delta_B-\phi)$, leading to a visibility:
\begin{equation}
V=\frac{2t_s^2t_l^2}{t_s^4+t_l^4}
\end{equation}
In our experiment, we typically obtain transmission differences
between the long and the short arm between 1 and 1.5 dB, which
limit the maximal visibility to around $(96 \pm 1)\%$. Moreover,
the states we create are not completely maximally entangled, due
to the fact that the first and the last pump pulses in a train
have a slightly smaller intensity. Finally, the interferometer
might not have a perfect visibility. To take into account these
last factors, we estimate a maximal visibility of $(99\pm 1)\%$.

Considering all the above mentioned factors, we find an optimal
visibility of $(92.2\pm 1.6 )\%$ which fits with the measured
value of $(91.6\pm 1.2 )\%$. This is a confirmation that the phase
noise is negligible and consequently that the coherence is kept
over many time-bins and that we generate entangled qudits.

In conclusion, we reported an experiment where we entangled two
photonic systems of arbitrary discrete dimensions. The simple
analysis presented in this paper already allows us to demonstrate
the creation of a photon pair in a coherent superposition of $d$
emission times, providing evidence of high-dimensional
entanglement. More complex analysis with d-arm interferometers
should allow to reveal all the quantum information content of such
states (e.g. nonlocality).

The authors would like to thank Claudio Barreiro and Jean-Daniel
Gautier for technical support. Financial support by the Swiss NCCR
Quantum Photonics, and by the European project RamboQ is
acknowledged.


\begin{references}
\bibitem{note0} This is an advantage compared to the multiphoton
schemes since the detection efficiency remains constant for any
dimension.
\bibitem{note1} The time difference between 2 pump pulses can be
determined very precisely by measuring the repetition rate of the
laser. This allows us to determine the path length difference of
the interferometer with a precision better than the pump coherence
length (i.e a few tens of microns)
\bibitem{note2}Each photon
can follow two paths \cite{note3}: long ($L$) or short($S$) arm of
the interferometer, so there are 4 different paths for the pair.
Only two of them ($[S,S]$ and $[L,L]$) lead to a coincident
detection.
\bibitem{note3} We restrict ourself to the
case where both photons exit the same output of the
interferometer. We omit to write a global factor
$(\frac{1}{2}t\eta)^2$, where $t$ is the transmission in the
interferometer and $\eta$ the quantum efficiency of the detectors.
But we take into account that $(t\eta)^2<<1$ when calculating
$R_2$:if a given "path" leads to several coincident detection, we
add the probabilities for each process.
\bibitem{note4}
Independently on how many pairs were present, we focus on the two
pairs that gave each one photon for the detection. The considered
event may have happened for these paths: $[L,x;y,L]$, $[S,x;y,S]$,
$[x,L;L,y]$ and $[x,S;S,y]$, where the left part corresponds to
the first pair and the right part to the second one, and
$x,y\in\{L,S\}$. Altogether, this gives 16 possible paths
\cite{note3}, and the total number of paths for four photons is
$4^2=16$.
\end{references}
\end{document}